\documentclass[aps,pra,twocolumn,groupedaddress,showpacs]{revtex4}

\usepackage{graphics}
\begin{document}

%\preprint{}

\title{Variance minimization variational Monte Carlo method}


\author{Imran Khan}
%\email[Email:]{ikhan@physics.utoledo.edu}
\author{Bo Gao}
\email[Email:]{bo.gao@utoledo.edu}
\homepage[Homepage:]{http://bgaowww.physics.utoledo.edu}
%\thanks{}
\affiliation{Department of Physics and Astronomy,
	University of Toledo, MS 111,
	Toledo, Ohio 43606}

%\altaffiliation{Permanent address: Department of Physics and Astronomy,
%	University of Toledo,
%	Toledo, Ohio 43606}

\date{\today}

\begin{abstract}

We present a variational Monte Carlo (VMC) method that works equally well for  
the ground and the excited states of a quantum system.  
The method is based on the minimization of the variance of energy, 
as opposed to the energy itself in standard methods.  
As a test, it is applied to the investigation of the 
universal spectrum at the van der Waals length scale for two identical Bose atoms 
in a symmetric harmonic trap, with results compared to the basically exact 
results obtained from a multiscale quantum-defect theory.

\end{abstract}

%\keywords{}

\maketitle

\section{Introduction}

Monte Carlo methods have played an important role in our understanding
of a variety of quantum systems, especially few- and many-body 
quantum systems with strong interactions that are difficult to 
treat otherwise (see, e.g., Refs.~\cite{kal74,cep95,fou01,gio99,blu01,dub01}). 
It is also well-known, however, that most quantum Monte Carlo 
methods \cite{cep95,fou01} are formulated in such a way that they are 
strictly applicable
only to the ground state of a quantum system, a restriction that has 
severely limited their applicability.
Consider, for example, the gaseous Bose-Einstein condensates (BEC) of 
alkali-metal atoms (see, e.g., \cite{dal99}).
Any theory that intends to treat the real atomic interaction 
has to deal with the fact that
the gaseous BEC branch of states are in fact highly excited states
of a many-atom system. There are many branches of states of lower
energies, including the first branch of liquid states as suggested and 
studied recently by one of us \cite{gao05b}. 

In this paper we present a variational Monte Carlo (VMC) method that works
the same way for either the ground or the excited states of a quantum 
system. It is based on the minimization of the variance 
of energy, and is the method underlying a recent investigation 
of the universal equation of state at the van der Waals 
length scale \cite{gao04a,gao05b} for few atoms in a trap \cite{kha06}.
The details of the method were skipped in the earlier article \cite{kha06},
both because the focus there was on a single gaseous BEC state,
which was not the best example illustrating the method,
and because there were no other independent results to directly compared 
with, except in the shape-independent limit \cite{kha06}. 

We present here, in Sec. II, the details of the variational Monte Carlo method 
based on the minimization of the variance of energy and shows that it 
applies equally well to the ground and the excited states of a quantum system.
In Sec. III, we present a better illustration of the method through the 
universal spectrum at the van der Waals length scale for two identical 
Bose atoms in a symmetric harmonic trap. It is an example where results for
multiple energy levels can be obtained independently using other 
methods \cite{bus98,tie00,blu02a,bol02}, including, in particular, 
a multiscale quantum-defect theory (QDT) \cite{che05,che07}.
Conclusions are given in Sec.~\ref{sec:conclusions}.
We point out that in the process of writing this article, 
we have discovered that an equivalent approach has been developed 
earlier by Umrigar \textit{et al.} \cite{umr88}. 
The derivation of our method, and the applications presented here
and earlier \cite{kha06}, are however different. 

\section{Variance Minimization variational Monte Carlo method}
\label{sec:vmvmc}

Consider the time-independent Schr\"{o}dinger equation
\begin{equation}
\widehat{H}\left|\Psi_n\right\rangle = E_n\left|\Psi_n\right\rangle \;,
\label{eq:sch}
\end{equation}
where the energy eigenstates $\left|\Psi_n\right\rangle$ form a complete,
orthonormal basis.

Existing quantum Monte Carlo methods are mostly based on the
fact that for an arbitrary trial wave function
satisfying proper boundary conditions, we have
\begin{equation}
E_T[\Psi_T]\equiv\frac{\left\langle\Psi_T \left|\widehat{H}\right|\Psi_T\right\rangle}
{\left\langle\Psi_T |\Psi_T\right\rangle}\ge E_0 \;,
\label{eq:minE}
\end{equation}
which means that the ground state wave function is the one that minimizes
the energy functional $E_T[\Psi_T]$. The proof can be found in standard 
quantum mechanics textbooks (see, e.g., \cite{sak94}).

The variance minimization variational Monte Carlo method (VMVMC), 
as proposed here, is based on the functional
\begin{equation}
\eta[\Psi_T] \equiv \frac{\left\langle\Psi_T \left|\widehat{H}^2\right|\Psi_T\right\rangle}
{\left\langle\Psi_T |\Psi_T\right\rangle} - 
\left[\frac{\left\langle \Psi_T \left|\widehat{H}\right|\Psi_T\right\rangle} 
{\left\langle\Psi_T |\Psi_T\right\rangle}\right]^2 \ge 0 \;.
\label{eq:vmvmc}
\end{equation}
The proof of Eq.~(\ref{eq:vmvmc}) and
its physical meaning can be best understood by expanding
the trial wave function using the complete basis defined by
Eq.~(\ref{eq:sch}) to write $\eta[\Psi_T]$ as
\begin{equation}
\eta[\Psi_T] = \frac{\sum_m\left|\left\langle\Psi_m|\Psi_T
	\right\rangle\right|^2(E_m-E_T)^2}
	{\sum_m\left|\left\langle\Psi_m|\Psi_T
	\right\rangle\right|^2}\;.
\label{eq:minDE}
\end{equation}
From Eq.~(\ref{eq:minDE}), it is clear that 
zero is the minimum of the functional $\eta[\Psi_T]$,
and this minimum is reached when and only when
$E_T=E_n$ and $\left\langle\Psi_m|\Psi_T\right\rangle=0$ for $m\neq n$,
namely, only when $|\Psi_T\rangle$ is an eigenstate of energy
as defined by Eq.~(\ref{eq:sch}).
This statement is equally applicable to the ground and the excited 
states of a quantum system.

The implementation of VMVMC, based on the minimization of 
the variance of energy $\eta[\Psi_T]$, is straightforward. 
It does not require much more than the standard VMC,
as we illustrate here using the example of identical
particles.

Consider $N$ identical particles in an external potential 
and interacting via pairwise interactions.
It is described by a Hamiltonian:
\begin{equation}
\widehat{H} = \sum_{i=1}^N\hat{h_i} + \sum_{i<j=1}^N v(r_{ij})\;,
\label{eq:ham1}
\end{equation}
with  
\begin{equation}
\hat{h_i} \ = \ -\frac{\hbar^2}{2m}\nabla_i^2 + V_{ext}(\mathbf{r}_i) \;.
\label{eq:ham2} 
\end{equation}
Here $V_{ext}(\mathbf{r})$ is the external ``trapping'' potential, 
and $v(r)$ is the interaction between particles.  

For the evaluation of the energy functional, we have
\begin{eqnarray}
\lefteqn{\langle\Psi_T |\widehat{H} | \Psi_T \rangle = 
\langle\Psi_T |N\hat{h_1} + \frac{1}{2}N(N-1)v_{12}|\Psi_T \rangle } \nonumber\\
	&=&  \int d\tau\Psi_T^*\Psi_T \frac{1}{\Psi_T}\lbrace N\hat{h_1} 
	+ \frac{1}{2}N(N-1)v(r_{12})\rbrace\Psi_T \nonumber\\
	  &=& \int d\tau\Psi_T^*\Psi_T E_{Loc}(\tau) \;,
\label{eq:ae}
\end{eqnarray}
where $\tau$ represents an $N$ particle configuration specified by their 
$3N$ coordinates. $E_{Loc}$ is the so-called local energy, and is given by
\begin{equation}
E_{Loc} = N\left( -\frac{\hbar^2}{2m}\right)\frac{1}{\Psi_T}\nabla_1^2\Psi_T 
	+ NV_{ext}(\mathbf{r}_1) + \frac{1}{2}N(N-1)v(r_{12})\;.
\end{equation}
The average energy is therefore
\begin{equation}
E_T = \frac{\int d\tau\Psi_T^*\Psi_T E_{Loc}(\tau)}{\int d\tau\Psi_T^*\Psi_T} \;.
\end{equation}
This is the standard integral in VMC, and can be evaluated using standard
Monte Carlo methods such the Metropolis method (see, e.g., \cite{thi99}).

In order to calculate the variance of energy, one must also determine 
the average of $\widehat{H}^2$.  
This can be done by first noting that, similar to Eq.~(\ref{eq:ae}),
we have
\begin{equation}
\left\langle \Psi_{m}|\widehat{H}|\Psi_T\right\rangle  
	=  \int d\tau\Psi_{m}^* \Psi_T E_{Loc}(\tau) \;,
\end{equation}
where $|\Psi_{m}\rangle$ is an eigenstate of energy as defined
by Eq.~(\ref{eq:sch}).
We have therefore
\begin{eqnarray}
\lefteqn{\left\langle \Psi_T\vert\widehat{H}^2\vert\Psi_T\right\rangle
= \sum_{m}\left\langle \Psi_T\vert\widehat{H}\vert\Psi_m\right\rangle  
	\left\langle \Psi_m\vert\widehat{H}\vert\Psi_T\right\rangle } \nonumber\\ 
&=& \sum_{m} \left\langle \Psi_m\vert\widehat{H}\vert\Psi_T\right\rangle^* 
	\left\langle \Psi_m\vert\widehat{H}\vert\Psi_T\right\rangle \nonumber\\
&=& \sum_{m}\int d\tau d\tau^\prime \left[
	\Psi_{m}( \tau^{\prime} ) \Psi_T^*(\tau^{\prime})
	E_{Loc}^*(\tau^{\prime}) \right. \nonumber\\
	& &\times\left.\Psi_{m}^*(\tau)\Psi_T(\tau)E_{Loc}(\tau) \right]\;.
\end{eqnarray}
Using the completeness relation
\begin{equation} 
\sum_{m} \Psi_{m}(\tau^{\prime})\Psi_{m}^*(\tau) = \delta(\tau^{\prime}-\tau) \;,
\end{equation}
we obtain
\begin{equation}
\left\langle \Psi_T\vert\widehat{H}^2\vert\Psi_T\right\rangle 
	= \int d\tau\Psi_T^*(\tau)\Psi_T(\tau)\vert E_{Loc}(\tau)\vert^2 \;,
\end{equation}
and therefore
\begin{equation}
\frac{\left\langle \Psi_T\vert\widehat{H}^2\vert\Psi_T\right\rangle}
	{\left\langle \Psi_T\vert\Psi_T\right\rangle}
	= \frac{\int d\tau\Psi_T^*(\tau)\Psi_T(\tau)\vert E_{Loc}(\tau)\vert^2}
	 {\int d\tau\Psi_T^*(\tau)\Psi_T(\tau)} \;.
\label{eq:ae2}	 
\end{equation}
The computation of the variance of energy, Eq.~(\ref{eq:vmvmc}),
has thus been reduced to two integrals, Eqs.~(\ref{eq:ae}) and (\ref{eq:ae2}),
both of which involving the same local energy, $E_{Loc}$,
that one encounters in standard VMC.
It is clear that the formulation and the equations in this section
are applicable to both bosons and fermions.

One can easily show that our method is equivalent to that of 
Umrigar \textit{et al.} \cite{umr88}.
However, we believe that our derivation provides a more rigorous
foundation and shows more explicitly why it works for both
the ground and the excited states.

\section{Sample Results for identical Bose Atoms in a symmetric harmonic trap}
\label{sec:results}

The VMVMC, as outlined in Sec.~\ref{sec:vmvmc}, was first applied
in Ref.~\cite{kha06} to study the universal equation of state
at the van der Waals length scale \cite{gao04a,gao05b}
for few identical Bose atoms ($N=3$-5) in a trap. 
To better illustrate and to further test the method, we investigate here
the universal spectrum at the van der Waals length scale 
for two identical Bose atoms in a symmetric harmonic trap. 
It is a problem for which accurate results can be obtained 
independently using a variety of methods \cite{bus98,tie00,blu02a,bol02}, 
including a multiscale QDT \cite{che05,che07}.

Two identical Bose atoms in a symmetric harmonic trap are described by the
Hamiltonian, Eqs.~(\ref{eq:ham1})-(\ref{eq:ham2}), with $N=2$, and
\begin{equation}
V_{ext}(\mathbf{r}_i) = \frac{1}{2}m\omega^2 r_i^2 \;,
\end{equation}
where $m$ is the mass of an atom, and $\omega$ is the trap frequency. 

For the trap states of interest here, we take the
trial wave function to be of the form of
\begin{equation}
\Psi_T = \left[\phi_{1}(\mathbf{r}_1)\phi_{2}(\mathbf{r}_2) + 
\phi_{1}(\mathbf{r}_2)\phi_{2}(\mathbf{r}_1)\right] F(r_{12})\;,
\label{eq:wfnt}
\end{equation}
where $\phi_1$ and $\phi_2$ are independent-particle orbitals,
and $F$ is the atom-atom correlation function that is
discussed in more detail in Ref.~\cite{kha06}. 
Specifically, we use
\begin{equation}
F(r) = \left\{ \begin{array}{lll} A u_\lambda(r)/r &,& r<d \\
			(r/d)^\gamma &,& r\ge d \end{array} \right. \;,
\label{eq:pcf}			 
\end{equation}
where $u(r)$ satisfies the Schr\"{o}dinger equation:
\begin{equation}
\left[-\frac{\hbar^2}{m}\frac{d^2}{dr^2} 
	+ v(r) - \lambda \right]
	u_{\lambda}(r) = 0 \;,
\label{eq:rsch}
\end{equation}
for $r<d$. $\gamma$ is the parameter characterizing the long-range
correlation between atoms in a trap, 
with $\gamma=0$ (meaning $F=1$ for $r>d$) corresponding to
no long-range correlation. 
Both $d$ and $\gamma$ are taken to be variational parameters, 
in addition to the variational parameters associated with
the descriptions of $\phi_1$ and $\phi_2$. 
The parameters $A$ and $\lambda$ are not independent.
They are determined by matching $F$ and its derivative at $d$.
Our choice of $F$ differs from traditional choices 
(see, e.g. Ref.~\cite{dub01}) not only in its treatment of the short-range
correlation, but especially in its allowance for the long-range 
correlation characterized by parameter $\gamma$. 
This was first suggested by a multiscale QDT treatment
of two atoms in a symmetric harmonic trap \cite{che05,che07},
and was later found to be the key for treating $N$ trapped
atoms in cases of strong coupling, namely when the $s$ wave 
scattering length $a_0$ becomes comparable to or greater than
the trap length scale $a_{ho}=(\hbar/m\omega)^{1/2}$ \cite{kha06}.

For atoms in their ground state, the atom-atom interaction
is of the van der Waals type of $-C_n/r^n$ with $n=6$
at large interatomic separations, i.e.,
\begin{equation}
v(r)\stackrel{r\rightarrow\infty}{\longrightarrow} 
	-C_6/r^6 \;.
\label{eq:vdW6}	
\end{equation}
This interaction has
an associated length scale of $\beta_6=(mC_6/\hbar^2)^{1/4}$,
and a corresponding energy scale of $s_E=(\hbar^2/m)(1/\beta_6)^2$ 
\cite{gao98a}.
Over a wide range of energies that is hundreds of $s_E$
around the threshold \cite{gao01,gao05a},
the details of atomic interactions of shorter range than 
$\beta_6$ are not important, and can be characterized by a 
single parameter that can be the $s$ wave scattering length
$a_0$, the short range $K$ matrix $K^c$, or some other related
parameters \cite{gao98b,gao01,gao04b}. In this range of energies, 
the spectrum of two atoms in a trap follows a 
universal property that can be characterized by \cite{kha06,che05,che07}
\begin{equation}
\frac{E_i/N}{\hbar\omega} = \Omega_i(a_{0}/a_{ho},\beta_6/a_{ho}) \;,
\label{eq:uspec}
\end{equation}
and is called the universal spectrum at length scale $\beta_6$.
Here $\Omega_i$ are universal functions that are uniquely determined by
the number of particles, the exponent of the van der Waals interaction 
($n=6$), and the exponent of the trapping potential (2 for the harmonic trap).
The strengths of interactions, characterized by $C_6$
and $\omega$, play a role only through scaling parameters such as
$\beta_6$ and $a_{ho}$. 

As in Ref.~\cite{kha06}, the universal spectrum at length scale
$\beta_6$, namely the $\Omega_i$'s in Eq.~(\ref{eq:uspec}), can be
computed by using a correlation function, Eq.~(\ref{eq:pcf}),
with $u_{\lambda}(r)$ as given by the angular-momentum-insensitive 
quantum-defect theory (AQDT) \cite{gao01},
\begin{equation}
u_{\lambda_s}(r_s) = B[f^{c(6)}_{\lambda_s l=0}(r_s) 
	- K^c g^{c(6)}_{\lambda_s l=0}(r_s)]\;.
\label{eq:wfn}
\end{equation}
Here $B$ is a normalization constant.
$f^{c(6)}_{\lambda_s l}$ and $g^{c(6)}_{\lambda_s l}$ are universal AQDT
reference functions for $-C_6/r^6$ type of potentials \cite{gao98a,gao04a}. 
They depend on $r$ only through a scaled radius $r_s=r/\beta_6$, 
and on energy only through a scaled energy $\lambda_s = \lambda/s_E$.
$K^c$ is the short-range K matrix \cite{gao01} that is related 
to the $s$ wave scattering length $a_{0}$ by \cite{gao03a,gao04b} 
\begin{equation}
a_{0}/\beta_n = \left[b^{2b}\frac{\Gamma(1-b)}{\Gamma(1+b)}\right]
	\frac{K^c + \tan(\pi b/2)}{K^c - \tan(\pi b/2)} \;,
\label{eq:a0s}
\end{equation}
where $b=1/(n-2)$, with $n=6$.
\begin{figure}
\scalebox{0.4}{\includegraphics{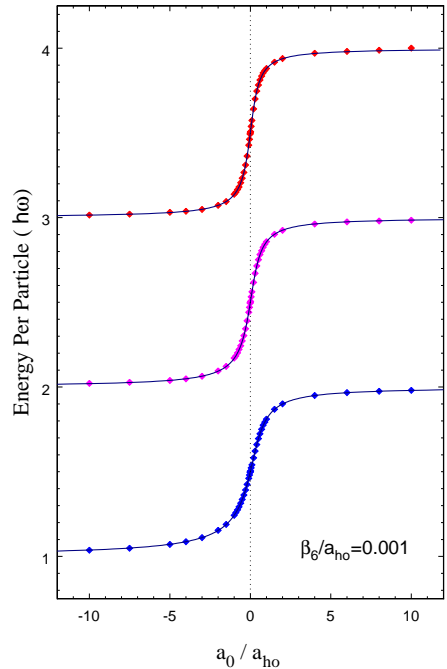} }
\caption{The universal spectrum at length scale $\beta_6$ 
for two Bose atoms in a symmetric harmonic trap 
as a function of $a_0/a_{ho}$ for $\beta_6 / a_{ho} = 0.001$.
Solid line: results from a multiscale QDT \cite{che07}. 
Symbols: results of VMVMC. \label{Figure1}}
\end{figure}
\begin{figure}
\scalebox{0.4}{\includegraphics{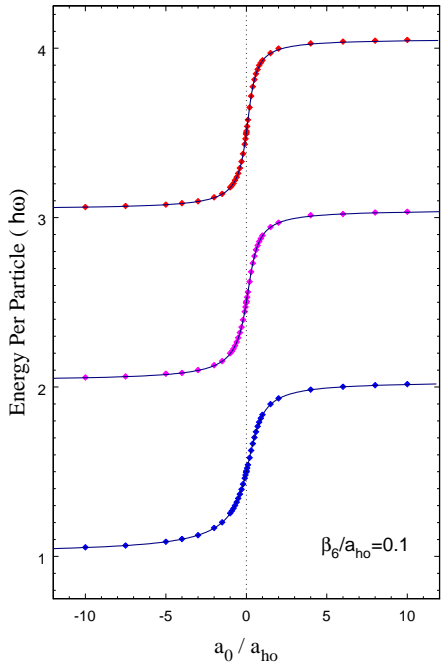} }
\caption{The same as Fig.~\ref{Figure1} except for $\beta_6 / a_{ho} = 0.1$.
\label{Figure2}}
\end{figure}

Figure~\ref{Figure1} shows a portion of the universal spectrum
at length scale $\beta_6$ for two Bose atoms in a symmetric harmonic
trap. Specifically, it gives the energies of the first three $s$ wave trap
states as a function of $a_0/a_{ho}$. The corresponding $\phi_i$s used 
in Eq.~(\ref{eq:wfnt}) are independent-particle orbitals
based on standard solutions for a single particle in a symmetric 
harmonic potential (see, e.g., \cite{zet01}).
For the lowest $s$ wave trap state, they are taken to be 
\begin{equation}
\phi_{i}(\textbf{r}) = exp(-\alpha_{i} x^2)\;, \;{i=1,2}.
\end{equation}
They are taken to be 
\begin{eqnarray}
\phi_{1}(\textbf{r}) &=& exp(-\alpha_{1} x^2) \;, \nonumber\\
\phi_{2}(\textbf{r}) &=& \left( \frac{3}{2} - x^2\right) exp(-\alpha_{2} x^2) \;,
\end{eqnarray}
for the first excited $s$ wave trap state, and
\begin{equation}
\phi_{i}(\textbf{r}) = \left( \frac{3}{2}-x^2\right) exp(-\alpha_{i} x^2)\;, \;{i=1,2},
\end{equation}
for the second excited $s$ wave trap state.
Here $x$ is a scaled radius defined by $x=r/a_{ho}$.
The variational parameters are $d$, $\gamma$, $\alpha_1$, and $\alpha_2$
in all three cases.
The variance of energy is calculated according to Sec.~\ref{sec:vmvmc},
and the minimization is carried out using a type of genetic algorithm.

\begin{figure}
\scalebox{0.4}{\includegraphics{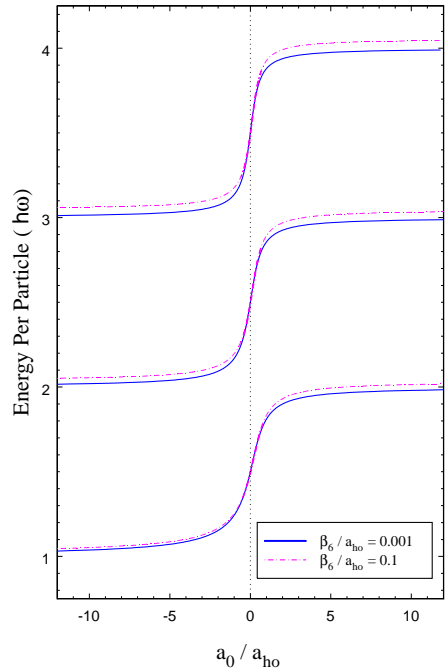} }
\caption{A comparison of the spectra for two different values of 
$\beta_6 / a_{ho}$, illustrating the shape-dependent correction 
that becomes more important for greater values of $\beta_6 / a_{ho}$
and for more highly excited states. \label{Figure3}}
\end{figure}

Both Figs.~\ref{Figure1} and \ref{Figure2} show that the results of VMVMC are in 
excellent agreements with those of a multiscale QDT \cite{che05,che07}, 
which gives basically exact results for two atoms in a symmetric harmonic trap.
(The scaled energy per particle, $E_i/(2\hbar\omega)$, used here is related to
the scaled center-of-mass energy, $e=\epsilon/\hbar\omega$, used in Ref.~\cite{che07},
by $E_i/(2\hbar\omega)=(e+3/2)/2$.)
The agreements are all within the variances of energy, which are smaller for
weaker coupling (smaller $a_0/a_{ho}$) and greater for stronger coupling,
but are in any case less than $1.8\times 10^{-3}$ for all parameters considered.
The results shown in Figure~\ref{Figure1}, which are for a small
$\beta_6 / a_{ho} = 0.001$, illustrate the shape-independent limit 
of $\beta_6 / a_{ho}\rightarrow 0$ for states with 
$E_i/2\sim\hbar\omega\ll s_E$ \cite{kha06,che07}. 
They agree, in this limit, with the results obtained using 
a delta-function pseudopotential \cite{bus98}.
For greater $\beta_6 / a_{ho}$, the effects of the van der Waals interaction
become gradually more important, especially for strong 
coupling ($a_0/a_{ho}\sim 1$ or greater) and for 
more highly excited states \cite{tie00,che07}. 
This is illustrated in Figure~\ref{Figure3}, which compares 
the results for $\beta_6 / a_{ho} = 0.1$ with those for 
$\beta_6 / a_{ho} = 0.001$.
We note that even the lowest trap state is itself a highly excited diatomic 
state. There are other ``molecular'' states that are lower in energy \cite{che05,che07}.
This fact does not, however, lead to any difficulties because VMVMC works
the same for the ground and the excited states.
It is for the same reason that we were able to investigate the 
gaseous BEC state for few atoms in a trap \cite{kha06}, 
which is again a highly excited state.
More detailed discussions of the universal spectrum at length scale
$\beta_6$ for two atoms in a symmetric harmonic trap, including 
the molecular states and the spectra for nonzero partial waves, 
can be found elsewhere \cite{che07}.

\section{Conclusions}
\label{sec:conclusions}

We have presented a variational Monte Carlo method, VMVMC, that works
the same for the excited states as it does for the ground state.  
The method is tested here through the universal spectrum at 
length scale $\beta_6$ for two identical Bose atoms in a symmetry harmonic trap, 
for which the results from VMVMC are found to be in excellent agreements 
with the basically exact results derived independently 
from a multiscale QDT \cite{che05,che07}.

\begin{acknowledgments}
This work was supported by the National Science Foundation under 
Grant No. PHY-0457060.
\end{acknowledgments}

\bibliography{sac,bose}

\end{document}